\documentclass[envcountsame,runningheads]{llncs}

\usepackage{graphics}
\usepackage{graphicx}
\usepackage[ruled,vlined,linesnumbered]{algorithm2e}

\usepackage{color}
\usepackage{xspace}
\usepackage{amsfonts}
\usepackage{amsmath}

\begin{document}
\title{A resource-frugal probabilistic dictionary and applications in (meta)genomics}

\author{Camille Marchet$^1$ \and Antoine Limasset$^1$ \and Lucie Bittner$^2$ \and Pierre Peterlongo\thanks{Corresponding author \email{pierre.peterlongo@inria.fr}}$^1$}
\institute{
$^1$ IRISA Inria Rennes Bretagne Atlantique, GenScale team\\
$^2$ Sorbonne Universit\'es, Universit\'e Pierre et Marie Curie (UPMC), CNRS, Institut de Biologie Paris-Seine (IBPS), Evolution Paris Seine, F-75005 Paris, France \\
}

\maketitle

\begin{abstract}
Genomic and metagenomic fields, generating huge sets of short genomic sequences, brought their own share of high performance problems.
To extract relevant pieces of information from the huge data sets generated by current sequencing techniques, one must rely on extremely scalable  methods and solutions. 
Indexing billions of objects is a task considered too expensive while being a fundamental need in this field.
In this paper we propose a straightforward indexing structure that scales to billions of element and we propose two direct applications in genomics and metagenomics.
We show that our proposal solves problem instances for which no other known solution scales-up. We believe that many tools and applications could benefit from either the fundamental data structure we provide or from the applications developed from this structure.
\end{abstract}


\section*{Introduction}
A genome or a chromosome can be seen as a word of millions characters long, written in a four letters (or \textit{bases}) alphabet. Modern molecular genome biology relies on sequencing, where the information contained in a genome is chopped into small sequences (around one hundred bases), called reads. By providing millions of short genomic reads along with reasonable sequencing costs, high-throughput sequencing technologies~\cite{schuster2007next} (HTS) introduced an era where data generation is no longer a bottleneck while data analysis is, as this amount of sequences needs to be pulled together in a coherent way.
Thanks to HTS improvements, it is possible to sequence hundreds of single genomes and RNA molecules, giving insight to diversity and expression of the genes. HTS even allowed to go beyond the study of an individual by sequencing all at once different species/organisms from the same environment, going from genomics to metagenomics. This massive sequencing represents a breakthrough: for instance one now can access and directly investigate the majority of the microbial world, which cannot be grown in the lab \cite{Hug2016}. However, because of the diversity and complexity of microbial communities, such experiments produce tremendous volumes of data, which represent a challenge for bioinformaticians to deal with.
The fragmented nature of genomic information, shredded in reads, craves algorithms to organize and make sense of the data.

A fundamental algorithmic need is to be able to index read sets for a fast information retrieval. In particular, given the amount of data an experiment can produce, methods that scale up to large data sets are needed. In this paper we propose a novel indexation method, called the quasi-dictionary, a probabilistic data structure based on Minimal Perfect Hash Functions (MPHF). This technique provides a way to associate any kind of data to any piece of sequence from a read set, scaling to very large (billions of elements) data sets, with a low and controlled false positive rate. 

A number of studies have focused on optimizing non-probabilistic text indexation, using for instance FM-index~\cite{Ferragina2000}, or hash tables. However, except the Bloomier filter~\cite{Charles2008}, to the best of our knowledge, no probabilistic dictionary has yet been proposed for which the false positive or wrong answer rates are mastered and limited. 
The quasi-dictionary mimics the Bloomier filter solution as it enables to associate a value to each element from a set, and to obtain the value of an element with a mastered false positive probability if the element was not indexed.  Existing published results in~\cite{Charles2008} indicate that the Bloomier filter and the quasi-dictionary have similar execution times, while our results tend to show that the quasi-dictionary uses approximately ten times less memory. 
Moreover, there are no available/free Bloomier filter yet implemented.


We propose two applications that use quasi-dictionary for indexing $k$-mers, enabling to scale up large (meta)genomic instances. As suggested by their names (\textit{short read connector counter} and \textit{short read connector linker}, as presented below), these applications have the ability to connect any read to either its estimated abundance in any read set or to a list of reads in any read set.
A  key point of these applications is to estimate read similarity using $k$-mers diversity only. This alignment-free approach is widely used and is a good estimation of similarity measure~\cite{Dubinkina2016}.

Our first application, called \textit{short read connector counter} (SRC\_counter), consists in estimating the number of occurrences of a read (i.e. its abundance) in a read set. This is a central point in high-throughput sequencing studies. Abundance is first very commonly used as indicator value for reads trimming: i.e. reads with relatively low abundance value are considered as amplification and/or sequencing errors, and these rare reads are generally removed before thorough analyses~\cite{kunin2010trim,Schirmer13012015}. The abundance of reads is then interpreted as a quantitative or semi-quantitative metric: i.e. reads abundance is used as a measure of genic or taxon abundance, themselves very commonly used for comparisons of community similarity~\cite{Amend2010,Kembel2012}.

The second proposed application in this work, called \textit{short read connector linker} (SRC\_linker), consists in providing a list of similar reads between sets. We define the read set similarity problem as follows. Given a read set \textit{bank} and a read set \textit{query}, provide a similarity measure between each pair of reads $b_i \times q_j$, with $b_i$ a read from the bank read set and $q_i$ a read from the query read set. Note that the bank and the query sets may refer to the same data set. Computing read similarity intra-read set or inter-read sets can be performed by a general purpose tool, such as those computing similarities using dynamic programming, and using heuristic tools such as BLAST~\cite{altschul1990basic}. 
However, comparing all versus all reads requires a quadratic number of read comparisons, leading to prohibitive computation time, as this is showed in our proposed results. There exists tools dedicated to the computation of distances between read sets~\cite{Benoit2016,maillet2014commet,Maillet2012}, but none of them can provide similarity between each pair of reads $b_i \times q_j$. Otherwise, some tools such as starcode~\cite{Zorita2015} are optimized for pairwise sequence comparisons with mainly the aim of clustering DNA barcodes. As shown in results, such tools also suffer from quadratic computation time complexity and thus do not scales-up data sets composed of numerous reads.

\paragraph{Availability and license} Our proposed tools SRC\_counter and SRC\_linker were developed using the GATB library~\cite{Drezen2014}. They may be used as stand alone tools or as libraries. They are licensed under the GNU Affero General Public License version 3 and can be downloaded from \url{http://github.com/GATB/rconnector}. Also licensed under the GNU Affero General Public License version 3, the quasi-dictionary can be downloaded from \url{http://github.com/pierrepeterlongo/quasi\_dictionary}.


\section{Methods}

We first recall needed notations. 
A $k$-mer is a word of length $k$ on an alphabet $\Sigma$. Given a read set $\mathcal{R}$, a $k$-mer is said \textit{solid} in $\mathcal{R}$ wrt a threshold $t$ if its number of occurrences in $\mathcal{R}$ is bigger or equal to $t$. Let $|w|$ denote the length of a word $w \in \Sigma^*$ and $|\mathcal{R}|$ denote the number of elements contained in a set $\mathcal{R}$.

The index we propose uses a minimal perfect hash function to associate each solid $k$-mer to a unique value in $[0,N-1]$, with $N$ being the total number of solid $k$-mers in $\mathcal{R}$. For an indexed $k$-mer, this value, called the $index$, can be used for downstream applications, as shown sections~\ref{vedere} and~\ref{soeur}.

Ideally, when querying a non indexed $k$-mer (i.e. a non solid $k$-mer or a $k$-mer absent from $\mathcal{R}$) the index returns no value. In our proposal, a non indexed $k$-mer may be associated to a value in $[0,N-1]$ with a probability $p>0$. We refer to our index as the  \textit{quasi-dictionary}, since it is a probabilistic index. However, note that querying any indexed $k$-mer provides a unique and deterministic answer.

In the following, we present our indexing solution, before to derive it to two metagenomic applications, in described sections~\ref{vedere} and~\ref{soeur}.

\subsection{Indexing solid $k$-mers using a quasi-dictionary}
\begin{figure}
\centering
    \begin{algorithm}[H]
    \SetAlgoNoEnd
    
    \KwData{ Read set $\mathcal{R}$, $k \in \mathbb{N}, t \in \mathbb{N}, f \in \mathbb{N}$}
    \KwResult{A quasi-dictionary QD}
    $k$-mer set $\mathcal{K}$ = get\_solid\_kmers$(\mathcal{R},k,t)$ \label{alg:qdcreate:createsolid}\;
    QD.MPHF = create\_MPHF$(\mathcal{K})$ \label{alg:qdcreate:createMPHF}\;
    \ForEach{$k$-mer $w$ in $\mathcal{K}$}{
        $index$ = QD.MPHF($w$)\;
        QD.FingerPrints[$index$] = create\_fingerprint($w$,$f$) \label{alg:qdcreate:createfingerprint}\;
        }
    \textbf{return} QD\;
    
    \caption{Quasi-dictionary creation. \label{alg:qdcreate}}
  \end{algorithm}
\end{figure}

Our quasi-dictionary proposal is designed for
the indexation of solid $k$-mers coming from a read set $\mathcal{R}$.
A quasi-dictionary is composed of two structures: a minimal perfect hash function MPHF (see for instance~\cite{Belazzougui2014}) and a table of fingerprints. 
Given a static set composed of $N$ distinct words, the MPHF constructs a structure that assigns to each word from the set a unique value in $[0,N-1]$. The fingerprint table is composed of $N$ elements. It assigns to each indexed word an integer value in $[O,2^f-1]$, with $f$ the size of the fingerprint in bits ($f\leq 2*k$ since a $k$-mer can be coded as a word of $2*k$ bits). As shown subsequently, this table is used to verify the membership of a query to the indexed set of words, with a false positive rate of $$\frac{2^{(2*k-f)}-1}{2^{2*k}}\approx\frac{1}{2^f}$$

Algorithm~\ref{alg:qdcreate} presents the construction of the quasi-dictionary. The set of solid $k$-mers (algorithm~\ref{alg:qdcreate}, line~\ref{alg:qdcreate:createsolid})  is obtained using the DSK~\cite{Rizk2013} method. The MPHF (algorithm~\ref{alg:qdcreate}, line~\ref{alg:qdcreate:createMPHF}) is computed using the MPHF library\footnote{\url{https://github.com/rizkg/BooPHF}, commit number $852cda2$}.

The fingerprint of a word $w$ (algorithm~\ref{alg:qdcreate}, line~\ref{alg:qdcreate:createfingerprint}) if obtained thanks to a hashing function $$ \text{create\_fingerprint: } \Sigma^{|w|} \rightarrow [0,2^f-1],$$ 
with $f \leq 2*k$. In practice we chose to use a xor-shift~\cite{marsaglia2003xorshift} for its efficiency in terms of throughput and hash distribution.

\begin{figure}
\centering
    \begin{algorithm}[H]
    \SetAlgoNoEnd
    
    \KwData{Quasi-dictionary QD, word $w$}
    \KwResult{A unique value in $[0,N-1]$ (with $N$ the number of indexed elements) or -1 if $w$ detected as non indexed}
    $index$ = QD.MPHF($w$)\label{alg:qdqueryMphf}\;
    
    \If{$index\geq0$ \textbf{and} QD.FingerPrints[$index$] = create\_fingerprint($w$)}{\textbf{return} $index$\;}\label{alg:qdqueryif}
   \textbf{return} $-1$\;
    
    \caption{Quasi-dictionary query \label{alg:qdquery}}
  \end{algorithm}
\end{figure}
The querying of a quasi-dictionary with a word $w$ is straightforward, as presented Algorithm~\ref{alg:qdquery}. The $index$ of $w$ is retrieved using the MPHF. Then the fingerprint stored for this $index$ is compared to the fingerprint of $w$. If they differ, then $w$ is not indexed and the $-1$ value is returned. If they are equal, the value $index \geq 0$ is returned. Note that two distinct words have the same fingerprint with a probability $\approx\frac{1}{2^f}$. It follows that there is a probability $\approx\frac{1}{2^f}$ that the quasi-dictionary returns a false positive value despite the fingerprint checking, \textit{i.e.} an $index\neq-1$ for a non indexed word. On the other hand, the $index$ returned for an indexed word is the correct one. 

In practice we use $f=12$ that limits the false positive rate to $\approx 0.02\%$. Note that our implementation authorizes any value $f \leq 64$. 
Moreover, depending on the MPHF implementation, non indexed words may be detected during the query (algorithm~\ref{alg:qdquery}, line~\ref{alg:qdqueryMphf}). We use an implementation in which more than $50\%$ of non indexed $k$-mers are attributed to the $-1$ $index$. Thus overall, the false positive rate of our proposal is limited to $\frac{1}{2*2^f}\approx 0.01\%$.


\paragraph{DNA strands}

As current sequencers usually do not provide the strand of each sequenced read, each indexed or queried $k$-mer should be considered in the forward or in the reverse complement strand. This is why, in the proposed implementations, we index and query only the canonical representation of each $k$-mer, which is the lexicographically smaller word between a $k$-mer and its reverse complement.

\paragraph{Time and memory complexities}

Our MPHF implementation has the following characteristics.
The structure can be constructed in $O(N)$ time and uses $\approx 4$  bits by elements. We could use parameters limiting memory fingerprint to less than 3 bits per element, but we chose parameters to speed up MPHF construction and query, and to be able not to return index for more than 50\% of non indexed elements, while this ratio is much smaller when using 3 bits per element ($\approx 33\%$).
The fingerprint table is constructed in $O(N)$ time, as the create\_fingerprint function runs in $O(1)$. This tables uses exactly $N\times f$ bits. Thus the overall quasi-dictionary size, with $f=12$ is $\approx 16$ bit per element. Note that this value does not take into account the size of the values associated to each indexed element.   

The querying of an element is performed in constant time and does not increases memory complexity.


\subsection{Approximating the number of occurrences of a read in a read set}\label{vedere}

\begin{figure}
\centering
\noindent\begin{minipage}{\textwidth}
\renewcommand\footnoterule{}                  
    \begin{algorithm}[H]
    \SetAlgoNoEnd
    \KwData{Read set $\mathcal{B}$, read set $\mathcal{Q}$, $k \in \mathbb{N}, t \in \mathbb{N}, f \in \mathbb{N}$}
    quasi-dictionary QD = create\_quasi-dictionary($\mathcal{B}$, $k,t,f$) \label{alg:count:qdcreate}\;
    create a table $count$ composed of $N$\footnote{with $N$ the number of solid $k$-mers from $\mathcal{B}$} integers\;
    \ForEach{Solid $k$-mer $w$ from $\mathcal{B}$}{
        $count[$query\_quasi-dictionary($w$)$]=$ number of occurrences of $w$ in $\mathcal{B}$\label{alg:count:set}\;
    }
    \ForEach{read $q$ in $\mathcal{Q}$}{
        create a empty vector $count\_q$\;
        \ForEach{$k$-mer $w$ in $q$ }{
            \If{query\_quasi-dictionary($w$) $\geq 0$ \label{alg:count:check}}{
                add $count[$query\_quasi-dictionary($w$)$]$ to $count\_q$ \label{alg:count:get}\;
            }
        }
        Output the $q$ identifier, and (mean, median, min and max values of $count\_q$)\;
    }
    \caption{SRC\_counter: Quasi-dictionary used for counting $k$-mers\label{alg:count}}
  \end{algorithm}
  \end{minipage}
\end{figure}

As presented in Algorithm~\ref{alg:count}, we propose a first straightforward application  using the quasi-dictionary. This application is called SRC\_counter for \textit{short read connector counter}. It approximates the number of occurrences of reads in a read set.

Two potentially equal read sets $\mathcal{B}$ and $\mathcal{Q}$ are considered. The indexation phase works as follows. Each solid $k$-mer of $\mathcal{B}$ is indexed using a quasi-dictionary. A third-party table named $count$ stores the counts of indexed $k$-mers. Elements of this table are accessed via the quasi-dictionary $index$ value of indexed items (Algorithm~\ref{alg:count} lines~\ref{alg:count:set} and~\ref{alg:count:get}). The number of occurrences of each solid $k$-mer from $\mathcal{B}$ (line~\ref{alg:count:set}) is obtained from DSK output, used during the quasi-dictionary creation (line~\ref{alg:count:qdcreate}). Then starts the query phase. Once the $count$ table is created, for each read $q$ from set $\mathcal{Q}$, the count of all its $k$-mers indexed in the quasi-dictionary are recovered and stored in a vector (lines~\ref{alg:count:check} and~\ref{alg:count:get}). Finally, collected counts from $k$-mers from $q$ are used to output an estimation of its abundance in read set $\mathcal{B}$. The abundance is approximated using the mean number of occurrences of $k$-mers from $q$, to supplement we output the median, the min and the max number of occurrences of $k$-mers from $q$. 

This algorithm is extremely simple. In addition to the quasi-dictionary creation time and memory complexities, it has a constant memory overhead (8 bits by element in our implementation) and it has an additional $O(\sum_{Q \in\mathcal{Q}}{|Q|})$ time complexity.

\subsection{Identifying similar reads between read sets or inside a read set}\label{soeur}

\begin{figure}[h]
\noindent\begin{minipage}{\textwidth}
\renewcommand\footnoterule{}                  
    \begin{algorithm}[H]
    \SetAlgoNoEnd
    \KwData{Read set $\mathcal{B}$, read set $\mathcal{Q}$, $k \in \mathbb{N}, t \in \mathbb{N}, f \in \mathbb{N}$}
    quasi-dictionary QD = create\_quasi-dictionary($\mathcal{B}$, $k,t,f$) \label{alg:link:qdcreate}\;
    create a table $ids$ composed of $N$\footnote{with $N$ the number of solid $k$-mers from $\mathcal{B}$}  vector of integers\label{alg:link:createtable}\;
    
    \ForEach{read $b$ in $\mathcal{B}$}{
        \ForEach{$k$-mer $w$ in $b$ }{
            $index$ = query\_quasi-dictionary($w$)\;
            \If{ $index \geq 0$ \label{alg:link:check}}{
                add id of $b$ to vector $ids[index]$ \label{alg:link:addid}\;
            }
        }
    }
    \ForEach{read $q$ in $\mathcal{Q}$}{\label{alg:link:startquery}
        create a hash table $targets$ ($target\_read\_id$) $\rightarrow$ couple($next\_free\_position$, $count$)\;
        \ForEach{i in $[0,|q|-k]$\footnote{In this work we consider sequence indices starting at 0}}{
            $w$ = $k$-mer occurring position $i$ in $q$\;
            $index$ = query\_quasi-dictionary($w$)\;
            
            \If{ $index \geq 0$ }{
                \ForEach{$target\_id$ in vector $ids[index]$}{
                    \If{$targets[target\_id]$ is empty}{
                        $targets[target\_id].next\_free\_position = i+k$\;
                        $targets[target\_id].count = 1)$\;
                    }
                    \Else{
                        \If{$i\geq targets[target\_id].next\_free\_position$}{\label{alg:link:testpos}
                            increase($targets[target\_id].count$)\label{alg:link:inccount}\;
                            $targets[target\_id].next\_free\_position = i+k$\label{alg:link:updatenextpos}\;
                        }
                    }
                }
            }
        }
        
        Output the id of $q$ and each\footnote{In practice only $target\_id$ whose $count$ value is higher or equal to a user defined threshold are output} $target\_id$ associate to its $count$ from $targets$ table\;
    }
    \caption{SRC\_linker: Quasi-dictionary used for identifying read similarities\label{alg:link}}
  \end{algorithm}
  \end{minipage}
\end{figure}

Our second proposal, called SRC\_linker for \textit{short read connector linker}, compares reads from two potentially identical read sets $\mathcal{B}$ and $\mathcal{Q}$. For each read $q$ from $\mathcal{Q}$, a similarity measure with reads from $\mathcal{B}$ is provided. 

The similarity measure we propose for a couple of reads $q\times b$ is the number of non-overlapping $k$-mers on $q$ that also occur on $b$. Note that this measure is not symmetrical as one does not verify that the $k$-mers do not overlap on $b$. 
Avoiding overlapping $k$-mers on $q$ enable to guarantee the span of the sequence on $q$ shared with $b$.

The indexation phase of SRC\_linker works as follows. A quasi-dictionary is created and a third-party table $ids$ of size $N$ is created. Each element of this table stores for a solid $k$-mers $w$ from $\mathcal{B}$  a vector containing the identifiers of reads from $\mathcal{B}$ in which $w$ occurs. See lines~\ref{alg:link:createtable} to~\ref{alg:link:addid} of Algorithm~\ref{alg:link}. 

The query phase (lines~\ref{alg:link:startquery} to the end of Algorithm~\ref{alg:link}) is straightforward, although a special care is taken to avoid overlapping on the query read $q$ in $\mathcal{Q}$. In practice, for each targeted read $b_j$ in $\mathcal{B}$ we remind the ending position of the last shared $k$-mer on $q$ with $b_j$ denoted by $next\_free\_position$  in the Algorithm~\ref{alg:link}. A new shared $k$-mer is counted (lines~\ref{alg:link:inccount} and~\ref{alg:link:updatenextpos}) only if its occurrence position in  $q$ is higher or equal to this position (line~\ref{alg:link:testpos}). 

Once all $k$-mers of a read $q$ are treated, the identifier of $q$ is output and for each read $b_j$ from $\mathcal{B}$ its identifier is output together with the number of shared $k$-mers with $q$. In practice, in order to avoid quadratic output size and for focusing on similar reads, only reads sharing a number of $k$-mers higher or equal to a user defined threshold are output. 

\paragraph{}
In addition to the quasi-dictionary data structure creation, considering a fixed read size, Algorithm~\ref{alg:link} has  $O(N\times \overline{m})$ memory complexity and a $O(N+\sum_{Q \in\mathcal{Q}}{|Q|\times \overline{m}})$ time complexity, with $\overline{m}$ the average number distinct reads from $\mathcal{B}$ in which a $k$-mer from $\mathcal{Q}$ occurs. In the worst case $\overline{m}=N$, for instance with $\mathcal{B}=\mathcal{Q}=\left\{A^{|\text{read}|}\right\}^N$. In practice, in our tests as well as for real set composed of hundred of million reads,  $\overline{m}$ is limited to $\approx 2.22$. 

\subsubsection*{Storing read identifiers on disk}

Storing the read identifiers as proposed in Algorithm~\ref{alg:link} presents important drawbacks as it requires a large amount of RAM memory. In order to get rid of this limitation we propose a disk version of this algorithm, in which the table $ids$ is stored on disk. As shown in Algorithm~\ref{alg:linkDisk} (Add. File), the algorithmic solution is not straightforward as one needs to know for each indexed $k$-mers $w$ its number of occurrences in the read set $\mathcal{B}$ plus the number of occurrences of $k$-mers $\neq w$ from $\mathcal{B}$ (false positives) that have the same quasi-dictionary $index$. 

This disk based solution enables to scale-up very large instances with frugal RAM memory needs, at the price of a longer computation time, as show in results. 

\section{Results}
This section presents results both about the fundamental quasi-dictionary data structure and about potential applications derived from its usage.
To this end, we use a metagenomic \textit{Tara} Oceans~\cite{Karsenti2011} read set ERR59928\footnote{ \url{http://www.ebi.ac.uk/ena/data/view/ERR599280}} composed of 189,207,003 reads of average size 97 nucleotides.  From this read set, we created six sub-sets by selecting first 10K, 100K, 1M, 10M, 50M and 100M reads (with K meaning thousand and M meaning million).

Tests were performed on a linux 20-CPU nodes running at 2.60 GHz with an overall of 252 GBytes memory.

\subsection{SRC\_counter tests and performances}

We provide SRC\_counter results enabling first to evaluate the gain of our proposed data structure when compared to a classical hash table. Secondly we provide results that enable to estimate the impact of false positives on results. 

\subsubsection{SRC\_counter performances compared to standard hash table index}

\begin{table}[h]
\centering
\begin{tabular}{c|c|cc|ccc|cc}
 \begin{tabular}{c}Indexed Dataset\\(nb solid $k$-mers)\end{tabular}& \begin{tabular}{c}
 $k$-mer count\\time (s)
 \end{tabular} &\multicolumn{2}{c|}{Construc. time (s)} &  \multicolumn{3}{c|}{Memory (GB)}&\multicolumn{2}{c}{Query Time(s)}\\
   & &QD  &Hash & QD & QD62 & Hash & QD  & Hash\\ \hline
 1M (64,321,167) & 2 &  34  & 106 & 0.25  &2.45  & 2.46 & 10 & 13\\ 
 10M (621,663,812) & 15 & 450 & 1091 & 1.80 & 5.45 & 23.58 & 11 & 17\\
 50M (2,812,637,134) & 72 &  2395 & 5027 & 8.00 & 16.37 & 106.25 & 11 & 19 \\
 100M (5,191,190,377) & 196 & 4855 & 9335 & 14.71 & 44.93 & 202.91 & 13 & 19\\
 Full (8,783,654,120) & 486 & 11671 &  & 24.83 & 75.96 & & 15 & \\
\end{tabular}
\caption{
Wallclock time and memory used by the SRC\_counter algorithm for creating and for querying the quasi-dictionary using the default fingerprint size $f=8$ (denoted by ``QD'') and the C++ \textit{unordered\_map}, denoted by ``Hash''. Column ``$k$-mer count time'' indicates the time DSK spent counting $k$-mers. 
Tests were performed using $k=31$ and $t=1$ (all $k$-mers are solid). The query read set was always the 10M set. We additionally provide memory results using the quasi-dictionary with a fingerprint size $f=62$ (denoted by ``QD62''). Construction and query time for QD62 are not shown as they are almost identical to the QD ones. On the full data set, using a classical hash table, the memory exceeded the maximal authorized machine limits (252 GB).}
\label{tab:qdbench}
\end{table}

We tested the SRC\_counter performances by indexing iteratively the six read subsets plus the full ERR59928 set, each time querying reads from set 10M. We compared our solution performances with a classical indexation scheme done using the C++ \textit{unordered\_map} hash table. Results are presented in Table~\ref{tab:qdbench}. These results show that the quasi-dictionary is  faster to compute  than a hash table solution. Moreover, the quasi-dictionary memory footprint is $\approx13$ times smaller on large enough instances (10 million indexed reads or more). Importantly these results show that the hash table is not a viable solution scaling up current read sets composed of several billions $k$-mers. Results also highlight the fact that the query is fast and only slightly depends on the number of indexed elements.

Importantly, using a fingerprint large enough (here $f=62$ for $k$-mers of length $k=31$), we can force the quasi-dictionary to avoid false positives. As expected, the quasi-dictionary data structure size increases with the size of $f$ but interestingly, on this example and as shown in Table~\ref{tab:qdbench}, the size of the quasi-dictionary with $f=62$ remains in average 4 times smaller than the size the hash-table on large problem instances. Keeping in mind that the quasi-dictionary is faster to construct and to query, the usage of this data structure avoiding false positives presents only advantages compared to the hash table usage for indexing a static set. However, one should recall that this is true because we are using an alphabet of size four, so any 31-mer on the alphabet $\{A,C,G,T\}$ can be assigned to a unique value in $[0,2^{62}-1]$ and \textit{vice versa}. With larger alphabets such as the amino-acids or the Latin ones, the usage of a hash table is recommended if false positives are not tolerated.

\subsubsection{Approximating false positives impact}
We propose an experiment to assess the impact on result quality when using a probabilistic data structure instead of a deterministic one for estimating read abundances. 

We used the read set 100M both for the indexation and for the querying, thus providing an estimation of the abundance of each read in its own read set. 
We made the indexation using $k=31, c=2$ and $f=8$. Note that, with $c=2$ only $k$-mers seen twice or more in the set are solid and thus are indexed. In this example only 756,804,245 $k$-mers are solid among the 5,191,190,377 distinct $k$-mers present in the read set. This means that during the query, 85.4\% of queried $k$-mers are not indexed. This enables to measure the impact of the quasi-dictionary false positives. 
We applied the count algorithm as described in Algorithm~\ref{alg:count}, and the tuned version using a hash table instead of a quasi-dictionary. We analyzed the count output composed of the average number of occurrences of $k$-mers of each read in the 100M read set.

Because of the quasi-dictionary false positives, results obtained using this structure are an over-estimation of the real result. Thus, we computed for each read, the observed difference in the counts between results obtained using the quasi-dictionary implementation and the hash table implementation. 
    The max over-approximation is 26.9, and the mean observed over-approximation is $7.27\times10^{-3}$ with a $3.59\times10^{-3}$ standard deviation. Thus, as the average estimated abundance of each read which  is $\approx 2.22$, the average count over-estimation represents $\approx0.033\%$ of this value. Such divergences are negligible. 

\subsection{Identifying similar reads}

We set a benchmark of our method with comparisons to state of the art tools that can be used in current pipelines for the read similarity identification presented in this paper.
We compared our tool with the classical method 
BLAST~\cite{altschul1990basic} (version 2.3.0), with default parameters. BLAST is able to index big data sets, and consumes a reasonable quantity of memory, but the throughput of the tool is relatively low and only small data sets were treated within the timeout (10h, wallclock time).
We also included two broadly used mappers in the comparison.
We used Bowtie2~\cite{langmead2012fast} (version 2.2.7), and BWA~\cite{li2009fast} (version 0.7.10) in any alignment mode (\texttt{-a} mode in Bowtie2, \texttt{-N} for BWA) in order to output all alignment found instead of the best ones only. Both tools are not well suited to index large set of short sequences nor to find all alignments and therefore use considerably more resources than their standard usage.

We also compared SRC\_linker to starcode (1.0), that clusters DNA sequences by finding all sequences pairs below a Levenshtein distance metric. One should notice that benchmark comparisons with tools as starcode is unfair as such tool provides much more precise distance information between pair of reads than SRC\_linker and performs additional task as clustering. However, our benchmark highlights the fact that such approaches suffer from intractable number of read comparisons, as demonstrated by presented results. 

\begin{table}[h]
\centering
\begin{tabular}{c|c|c|c|c|c||c|c|c|c|c|}
                 &\multicolumn{5}{c||}{Time(s)} &\multicolumn{5}{c|}{Memory(GB)}\\
 \begin{tabular}{c}
  Indexed\\Dataset
 \end{tabular} & Blast & Bowtie2  & BWA     & starcode & SRC\_linker & Blast & Bowtie2  & BWA       & starcode & SRC\_linker\\ \hline
 10K             &    4  &   3      &    6    & 2 &  1    & 0.7   &  0.29   &  0.04   & 11.36 & 1.01\\
 100K            &   52  &   51     &    106  & 29 &  5    & 18.5  &  0.77   & 0.49    & 12.06 & 1.07   \\
 1M              &   \textbf{795} & \textbf{10,644}   &  \textbf{3,155}  & \textbf{1,103} &  \textbf{45}   & \textbf{24.5}  &  \textbf{5.54}   & \textbf{3.4}    & \textbf{18.18} & \textbf{1.28} \\
 10M             &       &          &  62,912 & 131,139 & 587   &       &         & 5.9     & 73.5 &  3.61\\
 100M            &       &          &         &  & 14,748&       &         &         &  &  44.37\\
 Full            &       &          &         &  & 40,828&       &         &         &  &  110.84\\
\end{tabular}
\caption{CPU time and memory consumption for indexing and querying a data set versus itself. We set a timeout of 10h. BLAST crashed for 10M data set, Bowtie2 reached the timeout we set with more than 200h (CPU) for 10M reads. BWA performs best among the mappers, reaching the timeout for 100M reads (more than 200h (CPU) on this data set). On the 100M data set, starcode reached the timeout. Only SRC\_linker finished on all data sets. On the full data set, it lasted an order of magnitude comparable to what BWA performed on only 10M.}
\label{tab:mappersbench}
\end{table}

We focused on a practical use case for which our method could be used, namely retrieving similarities in a read set against itself. We used default SRC\_linker parameters ($k=31$, $f=12$, $c=2$).
Because of the limitations of the methods we used for the benchmark, reported in Table~\ref{tab:mappersbench}, we could compare against all methods only up to 1M reads. BWA performed better than the two other tools in terms of memory, being able to scale up to 10M reads, while Bowtie2 and BLAST could only reach 1M reads comparison. On this modest size of read set, we show that we are already ahead both in terms of memory and time. However the gap between our approach and others  increases with the amount of data to process. Dealing with the full Tara data set reveals the specificity of our approach (Table~\ref{tab:mappersbench}) that requires low resources in comparison to others and is able to deal with bigger data sets.

\begin{table}[h]
\centering
\begin{tabular}{c|cc|cc|c}
 &  \multicolumn{2}{|c|}{Indexation Time (s)} & \multicolumn{2}{|c|}{Query Time (s)} & Memory\\
 & \begin{tabular}{c}
  One\\
  thread
 \end{tabular}& \begin{tabular}{c}
  20\\
  threads\end{tabular} & \begin{tabular}{c}
  One\\
  thread
 \end{tabular}& \begin{tabular}{c}
  20\\
  threads\end{tabular}  & (GB)\\
 \hline
 
RAM Full & 18,067 & 1,768 & 17,558 & 992 & 110\\
Disk Full &  106,766 & 28,471 & 24,873 & 1,736 & 19\\

\end{tabular}
\caption{
Multithreading and disk performances.
The full read set was used to detail the performances of the RAM and Disk algorithm on a large data set.
We used default parameters $k=31$, $f=12$, $c=2$.  Times are  wallclock  times.}
\label{tab:Multithread}
\end{table}

Finally, we highlight that we provide a parallelised tool (10$\times$ speedup for the index and 17$\times$ speedup for query for RAM algorithm as shown in Table~\ref{tab:Multithread}) on the contrary to classical methods that are partly-parallelised as only the alignment step is well suited for parallelisation. The disk version does not fully benefit from multiple cores since the bottleneck is disk access. The main interest of this technique is a highly reduced memory usage at the price of an order of magnitude lower throughput, as presented Table~\ref{tab:Multithread}.

\section{Discussions and conclusion}

In this contribution, we propose a new indexation scheme based on a Minimal Perfect Hash Function (MPHF) together with a fingerprint value associated to each indexed element. 
This is a probabilistic data structure that has similar features than Bloomier filters, with smaller memory fingerprint.
This solution is resource-frugal (we have shown experiments on sets containing more than height billion elements indexed in $\approx 3$ hours and using less than 25GB RAM) and opens the way to new (meta)genomic applications. As proofs of concept, we proposed two novel applications: SRC\_counter and SRC\_linker. The first estimates the abundance of a sequence in a read set. The second detects similarities between pair of reads inter or intra-read sets. These applications are a start for broader uses and purposes.

Two main limitations of our proposal due to the nature of the data structure can be pointed out.
Firstly, compared to standard hash tables, our indexing data structure presents an important drawback: the exact set of keys to index has to be defined during the data structure creation and it has to be static. 
This may be a limitation for non fixed set of keys. Moreover, our data structure can generate false positives during query. Even with the proposed false positive ratio limited to $\approx 10^{-2}\%$ with defaults parameters, this may be incompatible with some applications. However we can force our tools to avoid false positives by using as a fingerprint the key itself. Interestingly, this still provides better time and memory performances than using a standard hash table in the DNA $k$-mer indexing context, with $k=31$, which is a very common value for read comparisons~\cite{Benoit2016}. 
Secondly, one should notice that our indexation proposal saves space regarding the association between an element and a specific array offset (if the element was indexed). However, our proposal does not limit the space needed for storing the value associated to each indexed element. Thus, with respect to classical hash tables, the memory gain is limited in problem instances in which large values are associated to each key. Indeed, in this case, the memory footprint is mainly due to the value over the indexing scheme. In order to benefit from our proposal even in such cases we proposed an application example in which the values are stored on disk. However, our approach is namely designed for problems where a huge number of elements to index are at stake, along with a small quantity of information to match with.

We could improve our technique to recognize key from the original set, using a technique from the hashing field~\cite{kirsch2006less} or from the set representation field~\cite{belazzougui2012compressed}.
In such framework, a set can be represented with less memory than the sum of the memory required by the keys. We could thus hope to be able to represent a non-probabilistic dictionary without storing keys. Otherwise, we could use the hashing information to achieve a smaller false positive rate with the same or a reduced memory usage. The main challenge will be to keep fast query operation for such complex data structure.

The results we provided show that alignment-based approaches do not scale when it comes to find similar reads in data sets composed of millions of sequences. The fact that HTS data count rarely less than millions reads justify our approach based on $k$-mer similarity. Moreover our approach is more straightforward and requires less parameters and heuristics than mapping approaches, that can sometimes turn them into blackboxes. However, such an approach remains less precise than mapping, since the $k$-mer order is not taken into account and is less sensitive because of the fixed size of $k$. An important future work will be to evaluate the differences between matches of our pseudo-alignment and matches of well-known and widely used tool as BLAST.

Our tools' property of enabling the test of a read set against itself opens the doors to applications such as read clustering. 
Latest sequencing technologies, called Third Generation Sequencers (TGS), provide longer reads~\cite{sharon2013single,tilgner2014defining} (more than a thousand bases instead of a few  hundreds for HTS). With previous HTS short reads, \textit{de novo} approaches to reconstruct DNA or RNA molecules were using assembly~\cite{grabherr2011full,robertson2010novo}, based on de Bruijn graphs. For RNA, these TGS long reads mean a change of paradigm as assembly is no more necessary, as one read is long enough to represent one full-length molecule. The important matter becomes to segregate families of RNA molecules within a read set, a purpose our approach could be designed for.

Furthermore, the methods we provide have straightforward applications examples in biology, such as the building of sequences similarity networks (SSN)~\cite{Atkinsonetal2009} using SRC\_linker. SSN are extremely useful for biologists because, in addition to allowing a user-friendly visualization of the genetic diversity from huge HTS data sets, they can be studied analytically and statistically using graph topology metrics. SSN have recently been adapted to address an increasing number of biological questions investigating both patterns and processes: e.g. population structuring~\cite{forsteretal2015,Fondi29042016}; genomes heterogeneity~\cite{boon2015}; microbial complexity and evolution~\cite{Corel2016}; microbiome adaptation~\cite{Bap2012Clinics,volkel:hal-01282715} or to explore the microbial dark matter \cite{dmSSN}. In metagenomic microbial studies, SSN offer an alternative to classical and potentially biased methods, and thus facilitate large-scale analyses and hypotheses generation, while notably including unknown/dark matter sequences sequences in the global analysis ~\cite{forsteretal2015,dmSSN}. Currently SSN are built upon general purposes tools such as BLAST. They thus hardly scale-up large data sets. A future work will consist in checking the feasibility of applying SRC\_linker for constructing SSN and, in case of success, to use it on large SSN problem instances on which other classical tools cannot be applied.


\section*{Acknowledgments}
This work was funded by French ANR-12-BS02-0008 Colib'read project. 
We thank the GenOuest BioInformatics Platform that provided the computing resources necessary for benchmarking. We warmly thank Guillaume Rizk and Rayan Chikhi for their work on the MPHF and for their feedback on the preliminary version of this manuscript. 
\bibliography{short_read_connectors}            

\section{Appendix}

Appendix contains a presentation of the SRC\_linker algorithm using disk for storing values (Algorithm~\ref{alg:linkDisk}).
\begin{figure}[h!]
\centering
\noindent\begin{minipage}{\textwidth}
\renewcommand\footnoterule{}                  
    \begin{algorithm}[H]
    \SetAlgoNoEnd
    \KwData{Read set $\mathcal{B}$, read set $\mathcal{Q}$, $k \in \mathbb{N}, t \in \mathbb{N}, f \in \mathbb{N}$}
    quasi-dictionary QD = create\_quasi-dictionary($\mathcal{B}$, $k,t,f$) \;
    create a table $ids$ composed of $N$\footnote{with $N$ the number of solid $k$-mers from $\mathcal{B}$} integers all valued to 0\;

    \ForEach{read $b$ in $\mathcal{B}$}{
        \ForEach{$k$-mer $w$ in $b$ }{
            $index$ = query\_quasi-dictionary($w$)\;
            \If{ $index \geq 0$}{
                add 1 to $ids[index]$\;
            }
        }
    }
    
     \ForEach{Solid $k$-mer $w$ from $\mathcal{B}$}{
           $index$ = query\_quasi-dictionary($w$)\;
           \If{ $index \geq 0$}{
                $count=ids[index]$\;
                $ids[index]=Temporary\_File.position$\;
                write $count+1$ '$0$' on $Temporary\_File$\;
            }
        }
        
     \ForEach{read $b$ in $\mathcal{B}$}{
        \ForEach{$k$-mer $w$ in $b$ }{
          $index$ = query\_quasi-dictionary($w$)\;
            \If{ $index \geq 0$}{
                $position=ids[index]$\;
                $Temporary\_File$.goto($position$)\;
                write id of $b$ in place of the first $0$ found\;
            }
        }
    }
    
    \ForEach{read $q$ in $\mathcal{Q}$}{
        create a hash table $targets$ ($target\_read\_id$) $\rightarrow$ couple($next\_free\_position$, $count$)\;
        \ForEach{i in $[0,|q|-k]$}{
            $w$ = $k$-mer occurring position $i$ in $q$\;
            $index$ = query\_quasi-dictionary($w$)\;
            \If{ $index \geq 0$ }{
                $position=ids[index]$\;
                 $Temporary\_File$.goto($position$)\;
                 read from  $Temporary\_File$ and put in vector $V$ all integer until a $0$ is found\;
                \ForEach{$target\_id$ in  $V$}{
                    \If{$targets[target\_id]$ is empty}{
                        $targets[target\_id].next\_free\_position = i+k$\;
                        $targets[target\_id].count = 1)$\;
                    }
                    \Else{
                        \If{$i\geq targets[target\_id].next\_free\_position$}{
                            increase($targets[target\_id].count$)\;
                            $targets[target\_id].next\_free\_position = i+k$\;
                        }
                    }
                }
            }
        }
        
        Output the id of $q$ and each\footnote{In practice only $target\_id$ whose $count$ value is higher or equal to a user defined threshold are output} $target\_id$ associate to its $count$ from $targets$ table\; 
    }
    
    \caption{SRC\_linker\_Disk: Quasi-dictionary used for identifying read similarities\label{alg:linkDisk}}
  \end{algorithm}
  \end{minipage}
\end{figure}

\end{document}